# Attention-Enriched Deep Learning Model for Breast Tumor Segmentation in Ultrasound Images

*Aleksandar Vakanski*[a*], *Min Xian*[a], and *Phoebe E. Freer*[b]
[a] Department of Computer Science, University of Idaho, Idaho Falls, USA
[b] University of Utah School of Medicine, Salt Lake City, USA
[*]Corresponding Author: Aleksandar Vakanski, 1776 Science Center Drive, Idaho Falls, ID 83402; Email, vakanski@uidaho.edu; Phone: (+1)208-757-5422

**Abstract**
Incorporating human domain knowledge for breast tumor diagnosis is challenging, since shape, boundary, curvature, intensity, or other common medical priors vary significantly across patients and cannot be employed. This work proposes a new approach for integrating visual saliency into a deep learning model for breast tumor segmentation in ultrasound images. Visual saliency refers to image maps containing regions that are more likely to attract radiologists' visual attention. The proposed approach introduces attention blocks into a U-Net architecture, and learns feature representations that prioritize spatial regions with high saliency levels. The validation results demonstrate increased accuracy for tumor segmentation relative to models without salient attention layers. The approach achieved a Dice similarity coefficient of 90.5% on a dataset of 510 images. The salient attention model has potential to enhance accuracy and robustness in processing medical images of other organs, by providing a means to incorporate task-specific knowledge into deep learning architectures.

**Introduction**
Computer-aided image analysis can assist radiologists' interpretation and diagnosis, and reduce error rates, as well as the level of stress regarding erroneous diagnosis (Cheng et al. 2010; Inoue et al. 2017; Jalalian et al. 2017; Moon et al. 2011; Wu et al. 2019). For instance, 3 to 6% of all radiologists' image interpretations contain clinically important errors, and also, significant variability in the inter- and intra-observer image interpretation is often reported (Elmore et al. 1994; Langlotz et al. 2019; Waite et al. 2016).

The emphasis in this work is on automated computer-aided diagnosis of tumors in breast ultrasound (BUS) images (Xian et al. 2018b). A large body of research work employed conventional and deep learning approaches to address tasks related to automated lesion localization, segmentation, and classification (Inoue et al. 2017; Jalalian et al. 2017; Litjens et al. 2017; Moon et al. 2011; Wu et al. 2019). In spite of this progress, existing methods lack robustness and consistency when processing images taken with different imaging equipment, where the variations in image intensity, contrast, and density often result in a degraded performance of models that otherwise perform well on custom-built datasets.

An important way to improve the performance of data-driven models is by incorporating prior domain-specific knowledge (Nosrati and Hamarneh 2016). On the other hand, incorporating prior knowledge in deep models for breast cancer detection is challenging, because unlike other medical organs—such as the kidney or the heart, whose features naturally lend themselves to the application of shape or boundary priors—breast tumors have a large variability in shape and boundaries from case to case. Extracting other priors in the form of curvature, texture, intensity, or number of regions for breast tumors is also not an option.

Our proposed approach incorporates topological and anatomical prior information into a deep learning model for image segmentation. More specifically, maps of visual saliency are employed for integrating image topology knowledge (Xu et al. 2016; Xu et al. 2018). The model for visual saliency estimation is formulated as a quadratic optimization problem, and it is based on calculations of neutro-connectedness between regions in the image (Xian et al. 2016; Xian 2017). Anatomical prior knowledge is integrated by decomposing the tissue layers into skin, fat, mammary, and muscle layers (Xu et al. 2019), and applying higher weights to the salient regions in images belonging to the mammary layer.

In this paper, we propose a novel approach to integrate domain knowledge into a deep neural network model by using the attention mechanism (Simonyan et al. 2013). A U-Net architecture (Ronneberger et al. 2015) is selected for incorporating the prior knowledge in the form of a pyramid of visual saliency maps. Attention blocks are integrated with the layers of the encoder to force the network to learn feature representations that place spatial attention to target regions with high saliency values. Unlike similar deep learning models that introduce attention blocks by merging internal feature representations from different layers (Chen et al. 2016; Jetley et al. 2018; Oktay et al. 2018b), the proposed approach employs external auxiliary inputs in the form of visual saliency maps for training the model parameters.

The main contributions of this paper are: (1) attention enriched deep learning model for integrating prior knowledge of tumor saliency; and (2) confidence level calculation for visual saliency maps.

The paper is organized as follows. The next section overviews related works in the literature. The Materials section describes the used image dataset. The Methods section covers the proposed network architecture, attention blocks, and visual saliency maps. Experimental validation is provided in the Results section. The Discussion section presents the findings of the experiments, and the Conclusion section summarizes the work.

**Related Works**

Computer-aided segmentation in medical imaging has been an important research topic for several decades, and it encompasses a vast body of work in the published literature. Recent advances in deep learning models (Goodfellow et al. 2016; LeCun et al. 2015) demonstrated great improvements in semantic image segmentation (Badrinarayanan et al. 2017; Chen et al. 2018a; Chen et al. 2018b; He et al. 2015; Lin et al. 2017; Long et al. 2015; Ronneberger et al. 2015; Zhao et al. 2017). Consequently, significant efforts have been devoted toward the implementation and design of deep neural networks for a wide range of medical applications, including segmentation of tumors and lesions (e.g., brain tumor (Kamnitsas et al. 2017), skin lesions (González-Díaz 2017), histopathology images (Chen et al. 2017; Graham et al. 2018; Kumar et al. 2017; Lin et al. 2018; Naylor et al. 2019)), and segmentation of organs (e.g., pancreas (Oktay et al. 2018b), lung (Hu et al. 2019), heart (Oktay et al. 2018a), or head and neck anatomy (Zhu et al. 2019)).

Likewise, the implementation of deep models for *breast tumor segmentation* has spurred interest in the research community in recent years (Xian et al. 2018b). Whereas the most popular image modality for this task have been ultrasound images (Abraham and Khan 2019; Chiang et al. 2019; Huang et al. 2018; Yap et al. 2018) and digital mammography images (Akselrod-Ballin et al. 2017; Dhungel et al. 2015; Jung et al. 2018; Kooi et al. 2017; Moor et al. 2018; Ribli et al. 2017), a body of literature used MRI (Jaeger et al. 2018), and histology images (Lin et al. 2018). U-Net (Ronneberger et al. 2015) and its numerous variants and modifications have been the most commonly used architecture for this problem to date. In spite of this progress, breast tumor segmentation is still an open research topic, due to challenges related to the inherent presence of noise and low contrast of images, sensitivity of current methods to the used image-acquisition method, equipment, and settings, and the lack of large open datasets of annotated images for training purposes.

*Priors in medical image segmentation.* Incorporating prior task-specific knowledge for medical image segmentation is important for improved model performance (Nosrati and Hamarneh 2016), and it can be crucial in tasks with small datasets of annotated medical images (i.e., most medical tasks at the present time). Prior knowledge can generally be in the form of shape, boundary, curvature, appearance (e.g., intensity, texture), topology (e.g., connectivity), anatomical information/atlas (structure of tissues or organs), user information (seed points or bounding boxes), moments (size, area, volume), distance (between organs and structures), and other forms. Although recent deep learning-based models have caused a leap of performance in image segmentation over conventional methods based on thresholding, region-growing, graph-based approaches, and deformable models (Cai and Wang 2013; Gómez-Flores and Ruiz-Ortega 2016; Huang et al. 2017; Liu et al. 2010; Rodrigues et al. 2015; Xiao et al. 2002), incorporating prior knowledge in deep neural networks has proven to be a difficult task, and consequently, has not been widely investigated. Namely, semantic image segmentation using deep networks typically relies on loss functions that optimize the model predictions at a pixel level, without taking into consideration inter-pixel interactions and semantic correlations among regions at the image level. To integrate prior knowledge into segmentation models, several works have proposed ***custom loss functions*** that enforce learning feature representations compatible with the priors. For instance, a loss function that penalizes both geometric priors (boundary smoothness) and topological priors (containment or exclusion of lumen in epithelium and stroma) was devised for histology gland segmentation (BenTaieb and Hamarneh 2016). Likewise, loss functions in fully convolutional networks (FCNs) that encode a shape prior were proposed for kidney segmentation (Ravishankar et al. 2017), cardiac segmentation (Oktay et al. 2018a), and segmentation of star shapes in skin lesions (Mirikharaji and Hamarneh 2018). The disadvantage of this approach is that the related models are task-specific and cannot be repurposed for segmentation of other objects of interest in medical images. Another line of research introduces a ***post-processing step*** with Conditional Random Fields where the segmentation predictions by a deep learning network are improved through assigning class labels to regions with similar topological properties (Chen et al. 2018a; Havaei et al. 2017; Huang et al. 2018). However, these methods increase the processing complexity and computational expense, and have been mostly replaced in recent years with end-to-end training models. Furthermore, a body of work proposed to incorporate shape priors by ***redesigning the network architecture***. For example, Li et al. (Li et al. 2016) employed an FCN with a VGG-16 base model where shape priors are learned by a consecutive concatenation of the original images with the obtained segmentation maps during several iterations of the procedure. Gonzalez-Diaz (González-Díaz 2017) created probability maps based on the knowledge of the patterns of skin lesions (e.g., dots, globules, streaks, or vascular structures) and merged them with extracted feature maps in a ResNet-based architecture. Furthermore, a boundary prior was incorporated in a deep learning model called deep contour-aware network (DCAN) that has two subnetworks for learning concurrently shapes and contour boundaries in histology images (Chen et al. 2017). Yet another class of methods utilizes ***generative models*** for introducing prior knowledge. E.g., in several early pre-FCN image segmentation models, Boltzmann machines networks were employed for learning shape priors (Chen et al. 2013; Eslami et al. 2014). A more recent research uses variational Bayes autoencoders for incorporating prior anatomical knowledge of the brain geometry in segmentation of MRI images (Dalca et al. 2018).

Despite the potential demonstrated by the above-described research work, to the best of our knowledge, there are no previous studies on the incorporation of prior knowledge in deep models for breast cancer detection. The challenge stems from the fact that unlike other medical organs (e.g., kidney, heart) where shape or boundary priors can be applied, such constraints are not applicable to breast cancer detection, due to the wide difference in the geometry of breast tumors. Analogously, it is difficult to extract generalized prior knowledge regarding curvature, moments, appearance, intensity, or number of regions for breast tumors. In this work, we introduce prior topology information in a deep learning segmentation model in the form of region connectivity and visual saliency. Such prior information is combined with anatomical prior knowledge of the tissue layers in breast images, as explained in the subsequent sections.

*Attention mechanism in deep learning*. Attention mechanism is an approach in deep networks layer design where the goal is to recognize discriminative features in the inner activation maps and to utilize this knowledge toward enhanced task-specific data representation and improved model performance (Simonyan et al. 2013). This mechanism contributes to suppressing less relevant features and emphasizing more important features for a considered task; e.g., in image classification, important features lie in salient spatial locations in the images.

Attention mechanism has been integrated into various deep learning models designed for image captioning (Li et al. 2018; Xu et al. 2015), language translation (Bahdanau et al. 2015), and image classification (Jetley et al. 2018; Wang et al. 2017). In general, attention in deep neural networks is traditionally implemented in two main forms, known as hard and soft attention. The implementation of hard (or stochastic) attention is non-differentiable, the training procedure is based on a sampling technique, and as a consequence, the models are difficult to optimize (Cao et al. 2015; Mnih et al. 2014; Stollenga et al. 2014). Soft (or deterministic) attention models are differentiable and trained with backpropagation; due to these properties, they have been the preferred form of implementation (Chen et al. 2016; Jaderberg et al. 2015; Wang et al. 2017). In image processing, the attention mechanism produces a probabilistic map of spatial locations in images, where the parameters of the attention map are learned in end-to-end training. Furthermore, the introduced architecture designs in image processing typically comprise of multiple attention maps with different resolutions, thereby capturing salient features across multiple levels of feature abstraction. For instance, Jetley et al. (Jetley et al. 2018) introduced attention gates at three intermediate layers in a VGG network, and a weighted combination of the attention maps is used in the last layer for image classification. Chen et al. (Chen et al. 2018a) introduce attention blocks in the initial DeepLab model for image segmentation, where attention weights are learned at different scales of a pyramidal feature representation.

Similar, attention gates were introduced in a U-Net architecture (Oktay et al. 2018b) and were employed in medical image processing for segmentation of the pancreas (Oktay et al. 2018b), and for breast tumor and skin lesion segmentation (Abraham and Khan 2019). This type of models uses the extracted features maps in the encoder path of the network for calculation of the attention maps, which are afterward merged with the up-sampled features maps in the decoder network, typically via element-wise multiplication. Such design forces the model to encode the locations and shapes of salient regions in extracted representations that are relevant for segmentation of the objects of interest. In the work by Tomita et al. (Tomita et al. 2018) an attention module was implemented in a 3D residual convolutional neural network to dynamically identify regions of interest (ROI) for processing high-resolution microscopy images, thus replacing the commonly used approach of sliding window ROI selection, and alleviating the computational burden in processing microscopy images. In a related work to the proposed approach, AttentionNet is designed on top of a ResNeXt encoder-decoder architecture and applies both spatial and channel attention blocks for segmentation of the anatomical tissue layers in BUS images (Li et al. 2019). Conversely to our method, the authors in (Li et al. 2019) did not apply AttenionNet for breast tumor detection, as well as they used activations maps of the intermediate layers of the network in the attention blocks.

**Materials and Methods**
The proposed approach is validated on a dataset of 510 breast ultrasound images (Xian et al. 2018a). The dataset is collected from three hospitals: the Second Affiliated Hospital of Harbin Medical University, the Affiliated Hospital of Qingdao University, and the Second Hospital of Hebei Medical University. All images in the dataset are de-identified, and informed consent to the protocol was obtained from all involved patients. Different types of imaging ultrasound devices were employed for acquiring the images, including GE VIVID 7 (General Electric Healthcare, Chicago, IL, USA), GE LOGIQ E9 (General Electric Healthcare, Chicago, IL, USA), Hitachi EUB-6500 (Hitachi Medical Systems, Chiyoda, Japan), Philips iU22 (Philips Healthcare, Amsterdam, Netherlands), and Siemens ACUSON S2000 (Siemens Healthineers Global, Munich, Germany). GE VIVID 7 and Hitachi EUB-6500 were used for collecting ultrasound images at Harbin Medical University, GE LOGIQ E9 and Philips iU22 were used at Qingdao University, and Siemens ACUSON S2000 was used at Hebei Medical University. Image annotation related to the segmentation and delineation of tumors in images was initially performed by three experienced radiologists, followed by voting and creating a single segmentation mask per image on which all three medical professionals agreed. Afterward, the annotations were reviewed by a senior radiologist expert, who either approved, or if needed, applied corrections and amendments to the segmentation boundaries (Xian et al. 2018a).

*Network Architecture*
The proposed network is based on the well-known U-Net architecture (Ronneberger et al. 2015), which consists of fully convolutional encoder and decoder sub-networks with skip connections. The layers in the encoder employ a cascade of convolutional and max-pooling layers, which reduce the resolution of input images and extract increasingly abstract features. The decoder comprises convolutional and up-sampling layers that provide an expanding path for recovering the spatial resolution of the extracted feature maps to the initial level of the input images. A unique characteristic of the U-Net architecture is the presence of skip connections from the feature maps in encoder's contracting path to the corresponding layers in the decoder. The features from the respective encoder's and decoder's layers are merged via concatenation that allows to recover the spatial accuracy of the objects in images and improves the resulting segmentation masks. Namely, although the central layer of the network offers high-level features with semantic rich data representation and a large receptive field, it also has low level of spatial context detail due to the down-sampling max-pooling layers along the contracting path, and impacts the localization accuracy around the object boundaries in the predictions. The skip connections provide a means to transmit low-level feature information from the initial high-resolution layers in the encoder to the reconstructing layers in the decoder, thereby restoring the local spatial information in predicted segmentations. Despite the introduction of deeper and more powerful models for image segmentation in recent years, the U-Net architecture has remained popular especially in medical image segmentation, where datasets have small size and large models can overfit on the available sets.

A graphical representation of the proposed model is presented in Figure 1. Besides the main input consisting of BUS images, the network has an auxiliary input consisting of the corresponding salient maps. Attention blocks introduce salient maps with reduced scale in all layers on the contracting path of the encoder in the form of an image pyramid. This enforces the network to focus the attention onto regions in the saliency maps with high intensity values. More specifically, the introduced attention blocks put more weights on areas in the extracted feature maps at each layer that have higher levels of saliency in the salient maps. Thus, the topology of the salient maps influences the learned feature representations.

The images and saliency maps are grey-scale 8-bit data resampled into floating point with normalization. Resized images and saliency maps to $256 \times 256$ pixels are used as inputs to the model. The number of convolutional filters per layer in the network is (32, 32, 64, 64, 128), which is reduced in comparison to the original U-Net, to account for the relatively small dataset. The output segmentation probability maps have the same spatial dimension as the inputs. The proposed network is trained in an end-to-end fashion; however, the saliency maps are precomputed and used at both training and inference.

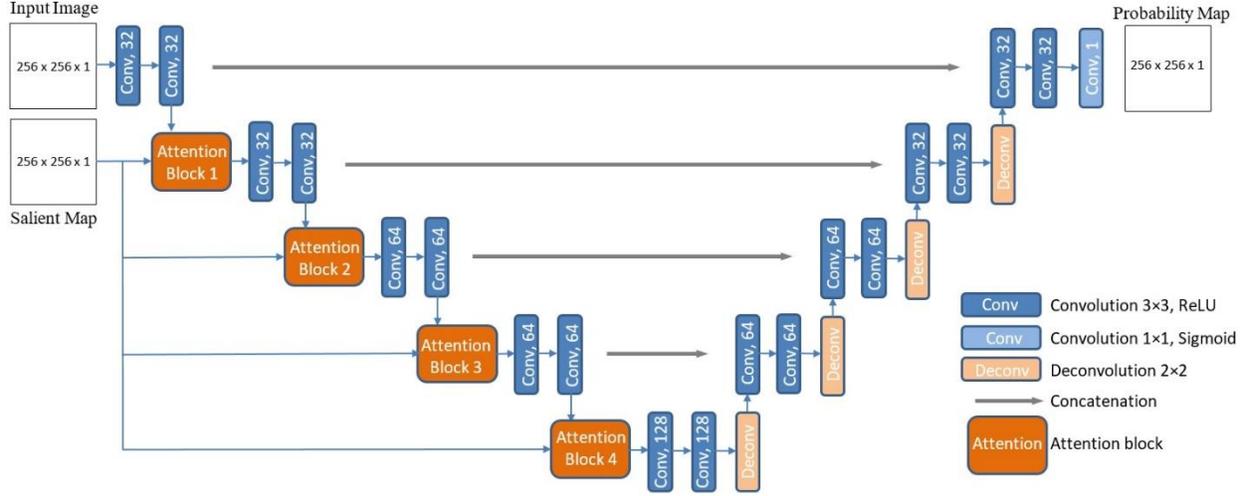

**Figure 1:** Architecture of the proposed U-Net model with salient attention. The model uses BUS images and saliency maps as inputs, and produces segmentation probability maps as outputs.

*Attention Blocks*

A block diagram of Attention Block *n* is depicted in Figure 2. The input feature maps to the attention block are denoted $F_n = \{f_1, f_2, \ldots, f_{k_n}\}$, where each feature map has horizontal and vertical spatial dimensions of $256/2^{(n-1)} \times 256/2^{(n-1)}$ pixels for the block in the layer level $n \in \{1, 2, 3, 4\}$. The symbol $k_n$ is the channel dimension of the feature maps in block $n$, i.e., $k_n \in \{32, 32, 64, 64\}$. For example, the input Feature Maps in Fig. 2 related to the output activations of the convolutional 'Conv 64' layer entering Attention Block 4 in Fig. 1 have dimensions of $32 \times 32 \times 64$ (i.e., for $n = 4$, the size of the feature maps $F_4$ is $256/2^3 \times 256/2^3 \times k_4 = 32 \times 32 \times 64$).

The input Salient Map in Fig. 2 is denoted $S$ and it is down-sampled through a max-pooling layer, resulting in $S_n$, which matches the spatial dimension of the input feature maps $F_n$ in Attention Block $n$. Next, $1 \times 1$ convolutions followed by rectified linear unit (ReLU) activation functions are used to increase the number of channels of the saliency map $S_n$ to 128. An element-wise sum block performs addition of $F_n$ and $S_n$ producing intermediate maps $I_n$ of size $256/2^n \times 256/2^n \times 128$. The intermediate maps $I_n$ are further refined through a series of linear $128 \times 3 \times 3$ and $1 \times 1 \times 1$ convolutions, followed by nonlinear ReLU activations. A sigmoid activation function normalizes the values of the activation maps into the [0, 1] range. The produced output is the attention map $A = (\alpha_i)$ with a spatial size of $256/2^n \times 256/2^n \times 1$, where the attention coefficients $\alpha_i$ have scalar values for each pixel $i$. Next, soft attention is applied via element-wise multiplication of the attention map $A$ with the max-pooled features $P_n$, i.e., $O_n = A * P_n$. The activation maps $O_n$ with size $256/2^n \times 256/2^n \times k_n$ are the Output of Attention Block $n$, and they are further propagated to the next layer, as depicted in Fig. 1.

The design of the attention block was inspired by the attention gates in (Oktay et al. 2018b) and (Jetley et al. 2018). Differently from these two works, where the attention blocks employ activation maps from the intermediate layers in the model as saliency maps for enhancing the discriminative characteristics of extracted intermediary features, the proposed attention block in this work utilizes precomputed saliency maps that point out to target spatial regions. If the attention block in this work applies directly the self-attention blocks described in (Abraham and Khan 2019; Jetley et al. 2018; Oktay et al. 2018b), the segmentation performance of the model would not improve. The reason for that lies in the distribution of salient regions in the used maps, since in many images background non-tumor areas have certain level of saliency in the salient maps; consequently, placing equal attention weights on all salient regions leads to higher level of false positive errors and degraded performance. The introduction of additional $3 \times 3$ and $1 \times 1$ convolutional layers for feature refinement in the proposed salient attention block was conducive toward improved segmentation outputs, which was confirmed via empirical validation of the proposed layers design.

*Saliency Maps*

Visual saliency estimation is an important paradigm for automatic tumor diagnosis in BUS images, where the aim is to model the level of saliency of image regions in correspondence to the capacity to attract radiologists' visual attention (Shao et al. 2015; Xie et al. 2017). For an input image, the output of such models is a visual saliency map with assigned saliency values in the [0, 1] range to every image pixel. High saliency value indicates a high probability that the pixel belongs to a tumor.

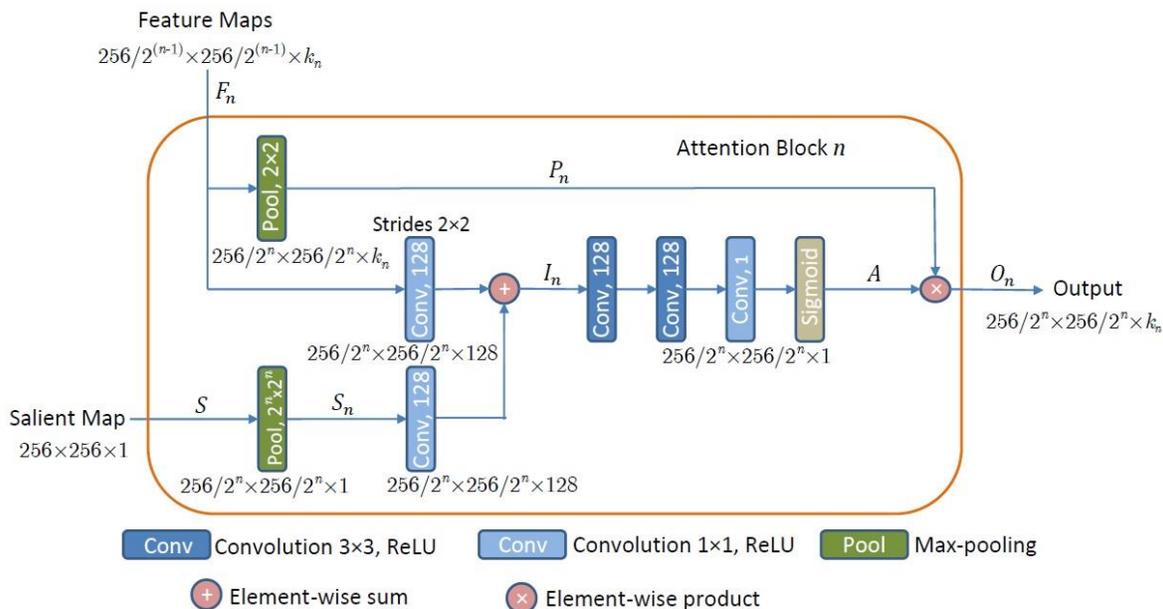

**Figure 2:** Attention block $n$, for $n \in \{1, 2, 3, 4\}$. Inputs to the block are feature maps from layer $n$ with spatial dimension $256/2^{(n-1)} \times 256/2^{(n-1)}$ with $k_n$ number of channels, and a salient map, and the output are down-sampled weighted maps with spatial dimension $256/2^n \times 256/2^n$ and $k_n$ number of channels.

The adopted approach for generating saliency maps of BUS images is based on our previous work (Xian et al. 2016; Xian 2017; Xu et al. 2016; Xu et al. 2018; Xu et al. 2019). In particular, the task of visual saliency estimation is formulated as a quadratic programming optimization that integrates high-level image information and low-level saliency assumptions. The model assigns a saliency value $s_i$ to each superpixel region $i$ in an image. The objective function of the model optimizes several terms, as follows. First, one term is a function of a foreground map that calculates the probability that the $i^{\text{th}}$ image region belongs to a tumor, and the distance between the $i^{\text{th}}$ region and the center of the foreground map of the image. Second, another term defines the cost of assigning zero saliency to an image region, and it employs the connectedness to the boundary regions to calculate the probability of the $i^{\text{th}}$ region belonging to a non-tumor image background. A third term applies a penalty if similar regions in the image have different saliency values. The formulation of the above functions is based on our Neutro-Connectedness (NC) approach (Xian, 2017; Xian et al., 2016) that exploits the information of the degree of connectedness and confidence of connectedness between the image regions. The complete set of formulas for derivation of the optimization model can be found in (Xu et al., 2019) and (Xu et al., 2018).

Our most recent work on this topic (Xu et al., 2019) introduces additional constraints in the model related to the breast anatomy by decomposing the images into four anatomical layers: skin, fat, mammary, and muscle layers. The four layers have different appearances in BUS images, and the fact that tumors are present predominantly in the mammary layer is used in our framework as an anatomical prior for saliency estimation. Two low-level saliency assumptions are utilized in the framework as well: 1) adaptive-center bias assumption forces the regions nearer the adaptive center to have higher saliency values; 2) the region-correlation assumption forces the similar regions to have similar saliency values. The extensive experiments in (Xu et al. 2019) showed the new model with anatomical knowledge generated improved performance than other models in related works on the dataset (Xian, 2018b). Another advantage of the approach proposed in (Xu et al. 2019) is the capability to interpret images without tumors, whereas many related approaches assume the presence of tumors in each image. Full implementation details can be found in the respective publications.

Examples of breast images and corresponding saliency maps are presented in Figure 3. The top row in the figure shows five BUS images, and the middle row displays the ground truth segmentation masks provided by radiologists. The bottom row displays the saliency maps for the images. One can note that the saliency maps assign a value to every pixel regarding the probability of belonging to a tumor, and differently from the ground truth masks, saliency values are assigned to background regions in images as well. Furthermore, the saliency maps are generated in an unsupervised manner, i.e., the information of the ground truth is not used by the saliency estimation model.

The incorporation of saliency maps into a deep learning model as complementary prior information is based on an assumption that the areas in images with high saliency values correspond to a high probability of tumor presence. Therefore, it is important that the saliency maps are of adequate quality and provide reliable information regarding the tumor locations. Otherwise, poor quality saliency maps can degrade the model performance.

The selected five examples of saliency maps depicted in Figure 3 have different levels of quality. More specifically, a map is considered of satisfactory quality when the location and intensity of the tumor region are clearly discernable in the saliency map. The example in the middle column in Figure 3 with moderate quality indicates the tumor location correctly, but the tumor shape and boundary do not match very well the ground truth, which may cause errors in the edge segmentation when applied to a deep network. For the case with low quality in Figure 3 there are several regions with similar area and saliency values, and it is not clear which of these regions may be tumors. Lastly, the saliency map with poor quality in Figure 3 assigns zero saliency values to the tumor region and completely misses the tumor.

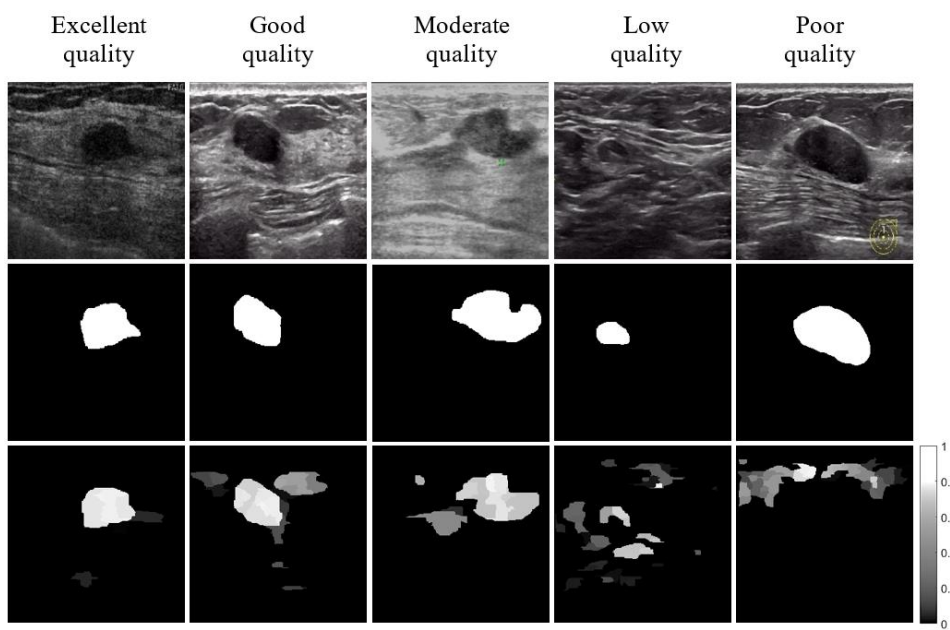

**Figure 3:** Examples of saliency maps with varying level of quality. Top row: original BUS image; Middle row: ground truth mask; Bottom row: saliency map.

In order to account for the cases with lower quality of saliency maps, we devised an algorithm that calculates the level of confidence in the saliency maps, and subsequently, eliminates the maps with low confidence. The approach is based on the following parameters: contour area $A_c = \sum_j p_j$ is the number of pixels of a contour $c$ in an image with a saliency value per pixel $p_j$ greater than a threshold value; cumulative intensity $I_c = \sum_j S(p_j)$ calculates the sum of the saliency values for the pixels in contour $c$; and, mean intensity $M_c = \sum_j S(p_j)/A_c$ of a contour $c$ is calculated as the ratio of the cumulative intensity and the area. The first rule in Algorithm 1 states that if the contour with the largest cumulative intensity $I_{argmax(I_{1:c})}$ has similar cumulative intensity to the second-largest contour and its mean intensity $M_{argmax(I_{1:c})}$ is not the highest of all contours (see Figure 4, left column), then eliminate the saliency map from the set. The second rule is similar to the first rule, and takes into account cases with larger ambiguities in the cumulative intensity and mean intensity of contours (Figure 4, middle column). The third rule considers the cases when a contour has high mean saliency intensity but smaller cumulative intensity than other contours in the image (see Figure 4, right column). The parameters in the algorithm are empirically set to: $a_1 = 2$, $a_2 = 3$, $a_3 = 0.2$, and $a_4 = 0.55$. In total 52 saliency maps satisfied the given conditions and were removed from the original set of 562 images, resulting in a reduced set of 510 images. That is, approximately 91% of the saliency maps are with high level of confidence. Having a low level of confidence for a saliency map does not necessarily mean that the saliency map is not correct: e.g., one can argue that the saliency for the example in the middle column in Figure 4 is correct. Rather, the proposed algorithm is designed to identify saliency maps with ambiguities regarding the spatial regions for tumor existence. The algorithm takes as inputs only the saliency maps, and it does not use the knowledge of the ground truth in estimating the level of confidence.

**Algorithm 1**: Confidence level calculation for saliency maps

For saliency map $i = 1:N$
    Find all fully connected contours with threshold $> 0.3$
    For contour $c = 1:C$
        if $I_{argmax(I_{1:c})} < a_1 I_{argmax(I_{1:c})-1}$ and $M_{argmax(I_{1:c})} < M_{argmax(M_{1:c})}$
        or if $I_{argmax(I_{1:c})} < a_2 I_{argmax(I_{1:c})-1}$ and $M_{argmax(I_{1:c})} + a_3 < M_{argmax(M_{1:c})}$
        or if $M_{argmax(M_{1:c})} > a_4$ and $argmax(M_{1:c}) \neq argmax(I_{1:c})$
        Remove saliency map $i$ from the set

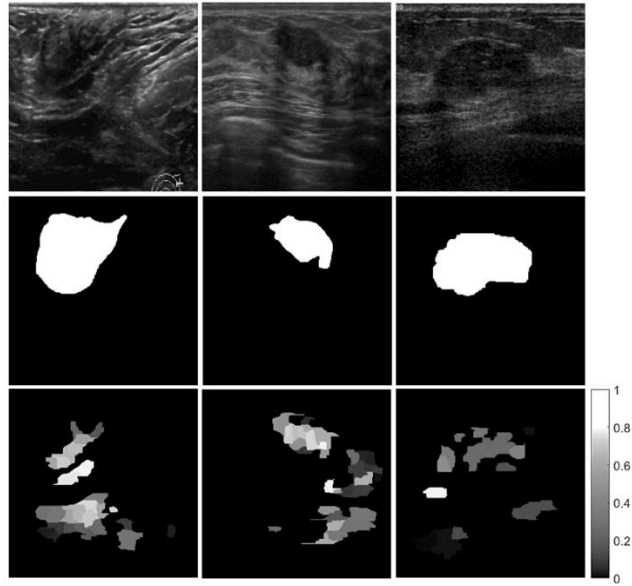

**Figure 4:** Examples of eliminated saliency maps from the original dataset. Top row: original BUS image; Middle row: ground truth mask; Bottom row: saliency map.

*Evaluation Metrics*
We used Dice similarity coefficient (DSC), Jaccard index (JI), true positives ratio (TPR), false positives ratio (FPR), and global accuracy (ACC) to evaluate the model performance:

$$DSC = \frac{2|A_g \cap A_p|}{|A_g| + |A_p|} \tag{1}$$

$$JI = \frac{|A_g \cap A_p|}{|A_g \cup A_p|} \tag{2}$$

$$TPR = \frac{|A_g \cap A_p|}{|A_g|} \tag{3}$$

$$FPR = \frac{|A_g \cup A_p - A_g|}{|A_g|} \tag{4}$$

$$ACC = \frac{|A_g \cap A_p| + |\overline{A_g} - A_g \cup A_p|}{|A_g| + |\overline{A_g}|} \tag{5}$$

In the above equations, $A_g$ is the set of pixels that belong to a tumor region in the ground truth segmented images, $\overline{A_g}$ is the set of pixels that belong to the background region without tumors in the ground truth segmented images, and $A_p$ is the corresponding set of pixels that are predicted to belong to a tumor region by the segmentation method. It is important to note that FPR is calculated as the ratio divided by the number of positives (i.e., pixels in tumor regions in the ground truth masks), as opposed to a ratio divided by the number of negatives (i.e., pixels in the background regions in the ground truth masks) as it is often defined in related tasks. Since the positive regions are smaller in BUS tumor segmentation, the selected formulation for FPR is more descriptive for this task. Additional metrics that we used for performance evaluation are the area under the curve of receiver operating characteristic score (AUC-ROC), Hausdorff distance (HD), and mean distance (MD). For most of the above metrics, the values are in the [0, 1] range, where higher values indicate improved performance (except for FPR, HD, and MD, where low values are preferred).

The differences in the values of the metrics obtained by different models are evaluated with a paired-comparison statistical hypothesis testing. A null hypothesis assumes that the metrics values are drawn from the same distribution and have a median value equal to zero.

*Implementation Details*

The proposed approach is validated on the described dataset of BUS images. We used five-fold cross-validation, where four folds (80% of images) are used for training, and one fold (20% of images) is used for testing. Validation during training is performed on 20% of the training set of images. All images in the dataset are first resized to a 256 × 256 pixels resolution. Since we focus on understanding the impact of the introduced salient attention on the model performance, we did not apply image augmentation.

The proposed model is trained with randomly initialized weights using Xavier normal initialization (Glorot and Bengio 2010). Dice loss function was used for training, defined as

$$\mathcal{L} = 1 - DSC = 1 - \frac{2|A_g \cap A_p|}{|A_g| + |A_p|} \tag{6}$$

where the same notation is preserved, i.e., $A_g$ and $A_p$ denote the ground truth and predicted masks, respectively.

The models were implemented using TensorFlow (Google, Menlo Park, CA, USA) and Keras (Francois Chollet, Menlo Park, CA, USA) libraries on the Google Colaboratory cloud computing services, which employ Tesla K80 GPUs. The network was trained by using adaptive moment estimation optimizer (Adam) with a learning rate of $10^{-4}$, and a batch size of 4 images. The training was stopped when the loss of the validation set did not improve for 20 epochs.

**Results**

*Evaluation and Comparative Analysis*

The experimental validation of the proposed approach is based on a comparative analysis of the following three models:

(1) U-Net;
(2) U-Net-SA. It applied the proposed salient attention approach; and
(3) U-Net-SA-C. It is a model with salient attention applied to a modified version where only one contour with the highest saliency is extracted in each salient map.

Examples of input BUS images, ground truth masks, saliency maps, and output segmentation maps by the models are presented in Figure 5. The values of the performance metrics are provided in Table 1. For the BUS images displayed in Figure 5, the segmentation outputs by the U-Net model are inferior in comparison to the predicted masks produced by the models with salient attention U-Net-SA and U-Net-SA-C. One particular aspect of improved performance entails the false positive predictions by U-Net (see rows A-G in Figure 5). In these cases, U-Net produces positive predictions of tumor presence for image regions that don't belong to a tumor. The attention models U-Net-SA and U-Net-SA-C benefited from the information in the salient maps, which led to a reduced rate of false positive predictions in A-G. This is especially noticeable in rows B, E, and G that have high quality salient maps, resulting in great improvement over the predictions by the basic U-Net model.

Furthermore, improved performance with respect to the true positive predictions by U-Net is displayed for rows H and I in Figure 5. The provision of salient maps for these two cases helps the model to focus on target regions with high saliency, leading to higher true positives rate of the segmentation masks by U-Net-SA over the basic U-Net model. In addition, rows J and K provide examples where the geometry of the salient regions in the saliency maps contributes to more accurate predictions of the proposed models in comparison to U-Net. Cases C and I are instances of BUS images with small size tumors, where the salient attention models successfully located the tumor regions. As explained earlier, the U-Net-SA-C model employs salient maps with one contour with the highest saliency intensity, and in many images it further improves the segmentation outputs. This is noticeable in row A in Figure 5, where the false positives in the segmentation are reduced in comparison to U-Net-SA. However, U-Net-SA-C model is based on an assumption that there is only one tumor in the images, which may not always be the case.

The results in Table 1 show the average and standard deviation (in parenthesis) per fold in the five-fold cross-validation procedure for the three deep models. The obtained values indicate that the models with salient attention U-Net-SA and U-Net-SA-C outperform the basic U-Net network without attention blocks for all performance metrics. The model U-Net-SA-C trained on the dataset with a single contour in the salient maps produced improved segmentation performance in comparison to U-Net-SA. The average training time per fold for the basic U-Net model was 7.58 minutes, whereas the corresponding times for training the salient attention models U-Net-SA and U-Net-SA-C were 8.54 and 8.08 minutes, respectively. Segmentation of the testing set of images with a trained model took 1.09, 1.26, and 1.37 seconds per fold (i.e., 102 images) for U-Net, U-Net-SA, and U-Net-SA-C, respectively. This translates to processing times of 12 milliseconds per image for U-Net-SA and 13 milliseconds per image for U-Net-SA-C.

**Table 1**: Performance evaluation metrics for models without and with salient attention. The shown values correspond to the average and standard deviation (in parenthesis) per fold in five-fold cross-validation.

| Model | DSC | JI (IOU) | TPR | FPR | ACC | AUC-ROC | HD | MD |
|---|---|---|---|---|---|---|---|---|
| **U-Net** | 0.894 (±0.013) | 0.821 (±0.017) | 0.903 (±0.011) | 0.107 (±0.019) | 0.978 (±0.002) | 0.951 (±0.006) | 4.346 (±1.377) | 0.224 (±0.240) |
| **U-Net-SA** | 0.901 (±0.013) | 0.832 (±0.014) | 0.904 (±0.016) | 0.092 (±0.008) | 0.979 (±0.001) | 0.955 (±0.002) | 4.326 (±1.360) | 0.209 (±0.234) |
| **U-Net-SA-C** | **0.905** (±0.013) | **0.838** (±0.014) | **0.910** (±0.011) | **0.089** (±0.012) | **0.980** (±0.001) | **0.957** (±0.004) | **4.271** (±1.326) | **0.201** (±0.218) |

A Wilcoxon signed rank test was adopted for statistical analysis, based on the distribution of the metrics values. The hypothesis testing results are presented in Table 2. The cells with asterisk indicate rejection of the null hypothesis with $P$-value $< 0.05$. ACC and AUC-ROC metrics are not included in the test since their values are calculated per a fold of 20% of the images, and not per individual images. Accordingly, for almost all metrics there is a statistically significant difference in the median values by the proposed models in comparison to U-Net. The exceptions are the TPR and HD values between U-Net and U-Net-SA, for which there isn't a statistically significant difference.

**Table 2**: Wilcoxon signed rank test of the performance metrics per image. * Statistically significant difference, $P$-value $< 0.05$.

| Model | DSC | JI (IOU) | TPR | FPR | HD | MD |
|---|---|---|---|---|---|---|
| **U-Net** and **U-Net-SA** | $P = 0.0011^*$ | $P < 0.0001^*$ | $P = 0.5822$ | $P < 0.001^*$ | $P = 0.2592$ | $P < 0.001^*$ |
| **U-Net** and **U-Net-SA-C** | $P < 0.0001^*$ | $P < 0.0001^*$ | $P = 0.0052^*$ | $P = 0.0098^*$ | $P = 0.0345^*$ | $P < 0.001^*$ |

Next, a comparison of our salient attention model for tumor segmentation U-Net-SA and three respective deep models for image segmentation is provided in Table 3. The dataset with 510 images is used for training the models. For a fair comparison, all models are trained in the same manner as the proposed architecture, i.e., five-fold cross-validation, batch size of 4, Xavier normal weights initialization, dice loss, Adam optimizer, and a stopping criterion of 20 epochs of non-improved validation loss. Due to the relatively small size of the dataset, for the comparison we selected smaller versions of the models. For instance, DenseNet is based on a network with 26 layers, and for PSPNet (that requires a base model) the small residual model ResNet18 is employed. The learning rate is fine-tuned for the different models, where an initial learning rate is selected, and when the validation loss does not improve for 10 epochs, the learning rate is reduced by a certain step size. The procedure is repeated until a preset value for the learning rate is reached. The details regarding the used learning rates for the different models are provided in Table 3. Our proposed U-Net-SA model listed last in the table outperformed the other deep learning networks for image segmentation on most of the employed performance metrics.

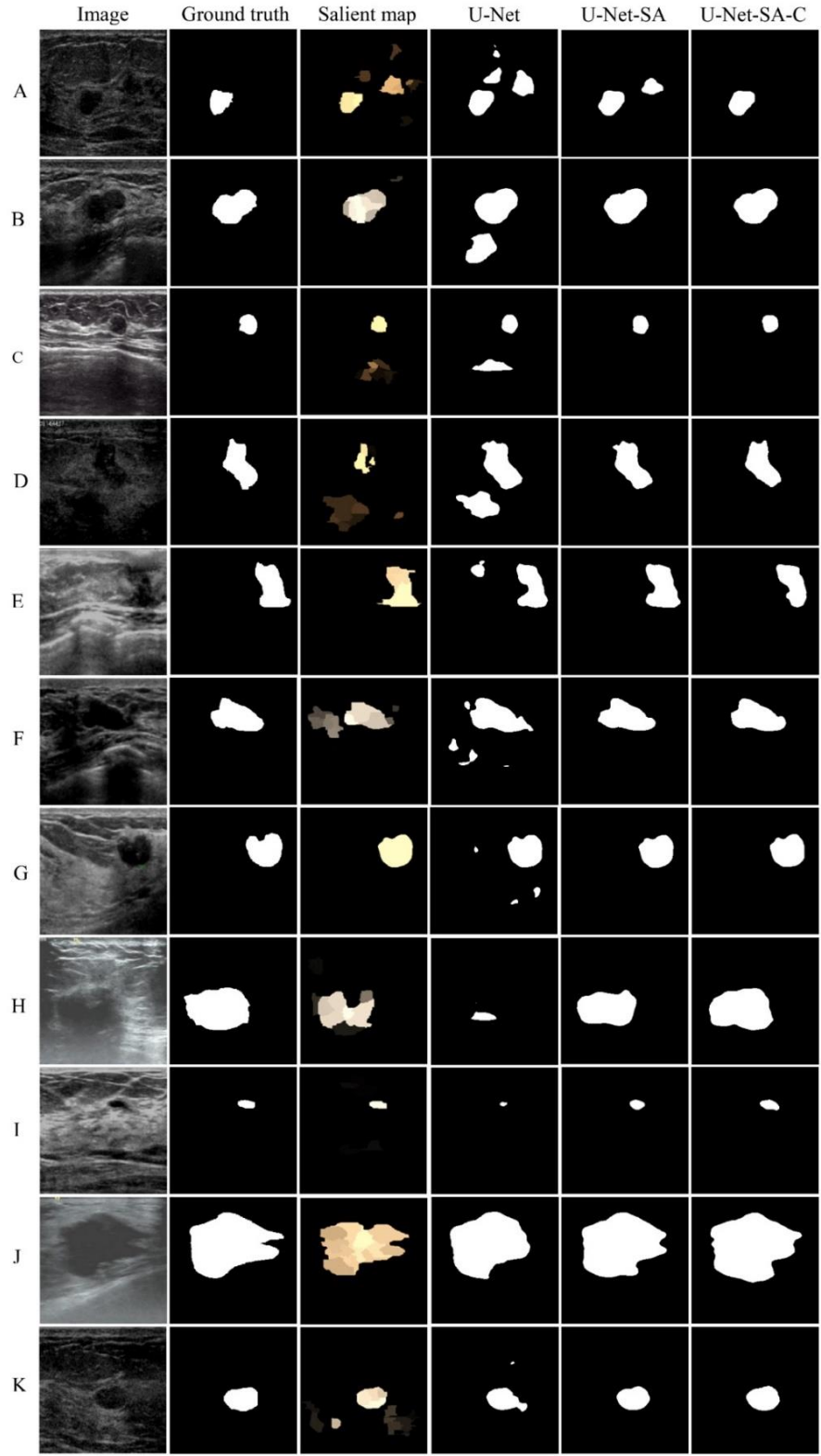

**Figure 5:** Segmentation results. First column: original BUS image; Second column: ground truth mask; Third column: saliency map; Fourth column: segmentation mask produced by U-Net; Fifth column: segmentation mask produced by U-Net-SA; Sixth column: segmentation mask produced by U-Net-SA-C.

**Table 3**: Values of the performance metrics for tumor segmentation by different models. The shown values correspond to the average and standard deviation (in parenthesis) per fold in five-fold cross-validation. LR represents the used learning rates for training the models.

| Model | Training Setting | DSC | JI (IOU) | TPR | FPR | ACC | AUC-ROC |
|---|---|---|---|---|---|---|---|
| **Seg-Net** | LR=$8 \cdot 10^{-4}$, decreased by 0.5 after 10 epochs until $1 \cdot 10^{-4}$ | 0.889 (±0.011) | 0.811 (±0.015) | 0.877 (±0.019) | **0.088** (±0.014) | 0.977 (±0.002) | 0.957 (±0.004) |
| **DenseNet-26** | LR=$1 \cdot 10^{-3}$, decreased by 0.1 after 10 epochs until $1 \cdot 10^{-4}$ | 0.888 (±0.016) | 0.818 (±0.017) | 0.886 (±0.019) | 0.093 (±0.025) | 0.978 (±0.002) | **0.958** (±0.005) |
| **PSPNet-ResNet18** | LR=$1 \cdot 10^{-4}$, decreased by 0.5 after 10 epochs until $5 \cdot 10^{-5}$, images size of 384x384 pix. | 0.886 (±0.008) | 0.808 (±0.008) | 0.884 (±0.014) | 0.107 (±0.016) | 0.976 (±0.002) | 0.953 (±0.005) |
| **Ours: U-Net-SA** | LR = $1 \cdot 10^{-4}$ | **0.901** (±0.013) | **0.832** (±0.014) | **0.904** (±0.016) | 0.092 (±0.008) | **0.979** (±0.001) | 0.955 (±0.002) |

Table 4 provides the values of the performance metrics for the models on the original dataset of 562 images. In comparison to the values presented in Table 1 on the reduced dataset of 510 images, the results in Table 4 indicate that the performances of the proposed attention enriched models U-Net-SA and U-NET-SA-C are reduced on the original dataset. Moreover, the basic U-Net model without salient attention has also reduced performance on the dataset of 562 images, which implies that the subset of 52 images that were removed from the original dataset contains breast tumors that are more challenging for segmentation in general. In conclusion, the algorithm for determining the level of confidence of the saliency maps contributed to improved performance on the reduced dataset of 510 images, by ensuring that the model predictions are not inhibited by poor data.

**Table 4:** Performance evaluation metrics for the models on the original dataset of 562 images. The shown values correspond to the average and standard deviation (in parenthesis) per fold in five-fold cross-validation. The values in bold font indicate the best performance for each metric.

| Model | DSC | JI (IOU) | TPR | FPR | ACC | AUC-ROC |
|---|---|---|---|---|---|---|
| **U-Net** | 0.891 (±0.005) | 0.817 (±0.008) | 0.900 (±0.009) | 0.120 (±0.027) | 0.977 (±0.002) | 0.950 (±0.006) |
| **U-Net-SA** | 0.894 (±0.006) | 0.824 (±0.008) | **0.901** (±0.017) | 0.111 (±0.032) | 0.978 (±0.002) | 0.952 (±0.012) |
| **U-Net-SA-C** | **0.896** (±0.007) | **0.825** (±0.010) | 0.899 (±0.020) | **0.106** (±0.025) | 0.978 (±0.002) | **0.955** (±0.010) |

**Discussion**

Based on the evaluation results presented in Table 1, the models with attention blocks outperformed the basic U-Net model. In addition, if only one contour with the highest saliency is extracted in the saliency maps (the U-Net-SA-C model), the performance improves further. This can be explained by the increasing spatial attention to a single salient region in the maps, resulting in reduced false positives in the outputs. As we mentioned earlier, this is based on an assumption that there is only one tumor in the images, which may not always be the case.

The design of the attention blocks has an impact on the segmentation output; therefore, we investigated several alternatives for the block layers and their parameters. Compared to similar attention blocks in deep models (Chen et al. 2016; Jetley et al. 2018; Oktay et al. 2018b), the used block in this work requires additional feature refinement by using convolutional 3×3 and 1×1 layers. The refinement layers balance the impact of inaccurate boundaries of the regions in salient maps on the learned features. In other words, the saliency maps do not provide accurate local information of the edges and boundaries of tumors in images, but rather, they provide global information of the spatial probability regarding the presence of tumors. Larger values of the attention coefficients put more emphasis on the edges and boundaries in salient maps and can reduce the segmentation outputs. The use of additional refinement layers lessens the values of the attention coefficients and results in improved tumor segmentation.

The fact that the ultrasound images for validation of the approach were collected with various imaging systems is a strength of the paper, as it makes the dataset suitable for training data-driven models with enhanced robustness to variations across images from different sources.

One limitation of the presented approach is that it relies on the quality of saliency maps. Using low quality maps can at best not improve the results, or result in degraded performance. To deal with this shortcoming, we proposed an algorithm that calculates a confidence score and eliminates the saliency maps with low confidence in their level of quality. Whereas visual saliency estimation is not the focus of this work, improvements in the models for visual saliency estimation can lead to improved segmentation by the proposed approach.

Avenues for future work include investigation of custom loss functions in deep learning models for encoding prior information, and working with medical partners to obtain annotated images with breast tissue layers and afterward integrating such anatomical prior with salient maps in a unified segmentation model.

**Conclusion**

This paper proposes a novel deep learning architecture that incorporates radiologists' visual attention for breast tumor segmentation. The proposed architecture consists of a variant of the basic U-Net model with attention blocks integrated along the contracting path in the layers of the encoder. The proposed attention blocks allow the deep learning model to suppress spatial regions with low saliency values, and respectively, to focus on regions with high saliency values. The attention blocks use multi-scaled versions of the saliency maps. The approach is validated on a dataset of 510 images, and the results demonstrate improved segmentation performance. The importance of this work stems from the difficulties in incorporating priors into deep learning models for medical image processing, and in particular for segmentation of breast ultrasound images, where most of the traditionally used prior forms cannot be applied.

**Acknowledgments**

This work was supported by the Center for Modeling Complex Interactions (CMCI) at the University of Idaho through NIH Award #P20GM104420. We would like to thank Fei Xu for providing the visual saliency maps from her latest research and for her constructive feedback and review of the paper.